\documentclass[twocolumn,showpacs,preprintnumbers,amsmath,amssymb,superscriptaddress]{revtex4}

\usepackage{graphicx}
\begin{document}
\title{Photoemission from buried interfaces in SrTiO$_{3}$/LaTiO$_{3}$ superlattices}

\author{M.~Takizawa}
\affiliation{Department of Physics and Department of Complexity 
Science and Engineering, University of Tokyo, 
5-1-5 Kashiwanoha, Kashiwashi, Chiba, 277-8561, Japan}
\author{H.~Wadati}
\affiliation{Department of Physics and Department of Complexity 
Science and Engineering, University of Tokyo, 
5-1-5 Kashiwanoha, Kashiwashi, Chiba, 277-8561, Japan}
\author{K.~Tanaka}
\affiliation{Department of Physics and Department of Complexity 
Science and Engineering, University of Tokyo, 
5-1-5 Kashiwanoha, Kashiwashi, Chiba, 277-8561, Japan}
\author{M.~Hashimoto}
\affiliation{Department of Physics and Department of Complexity 
Science and Engineering, University of Tokyo, 
5-1-5 Kashiwanoha, Kashiwashi, Chiba, 277-8561, Japan}
\author{T.~Yoshida}
\affiliation{Department of Physics and Department of Complexity 
Science and Engineering, University of Tokyo, 
5-1-5 Kashiwanoha, Kashiwashi, Chiba, 277-8561, Japan}
\author{A.~Fujimori}
\affiliation{Department of Physics and Department of Complexity 
Science and Engineering, University of Tokyo, 
5-1-5 Kashiwanoha, Kashiwashi, Chiba, 277-8561, Japan}
\author{A.~Chikamatsu}
\affiliation{Department of Applied Chemistry, University of Tokyo, 
Bunkyo-ku, Tokyo 113-8656, Japan}
\author{H.~Kumigashira}
\affiliation{Department of Applied Chemistry, University of Tokyo, 
Bunkyo-ku, Tokyo 113-8656, Japan}
\author{M.~Oshima}
\affiliation{Department of Applied Chemistry, University of Tokyo, 
Bunkyo-ku, Tokyo 113-8656, Japan}
\author{K.~Shibuya}
\affiliation{Institute for Solid State Physics, University of Tokyo, 5-1-5 Kashiwanoha, Kashiwashi, Chiba, 277-8581, Japan}
\author{T.~Mihara}
\affiliation{Materials and Structures Laboratory, Tokyo Institute of Technology, 4259 Nagatsuta, Midori-ku, Yokohama, Kanagawa, 226-8503, Japan}
\author{T.~Ohnishi}
\affiliation{Institute for Solid State Physics, University of Tokyo, 5-1-5 Kashiwanoha, Kashiwashi, Chiba, 277-8581, Japan}
\author{M.~Lippmaa}
\affiliation{Institute for Solid State Physics, University of Tokyo, 5-1-5 Kashiwanoha, Kashiwashi, Chiba, 277-8581, Japan}
\author{M.~Kawasaki}
\affiliation{Institute for Materials Research, Tohoku University, 
  2-1-1 Katahira, Aoba-ku, Sendai, Miyagi, 980-8577, Japan}
\author{H.~Koinuma}
\affiliation{Materials and Structures Laboratory, Tokyo Institute of Technology, 4259 Nagatsuta, Midori-ku, Yokohama, Kanagawa, 226-8503, Japan}
\author{S.~Okamoto}
\affiliation{Department of Physics, Columbia University, 538 West 120th Street, New York, New York 10027, USA}
\author{A.~J.~Millis}
\affiliation{Department of Physics, Columbia University, 538 West 120th Street, New York, New York 10027, USA}
\date{\today}

\begin{abstract}
We have measured photoemission spectra of SrTiO$_{3}$/LaTiO$_{3}$ superlattices with a topmost SrTiO$_{3}$ layer of variable thickness. Finite coherent spectral weight with a clear Fermi cut-off was observed at chemically abrupt SrTiO$_{3}$/LaTiO$_{3}$ interfaces, indicating that an ``electronic reconstruction'' occurs at the interface between the Mott insulator LaTiO$_{3}$ and the band insulator SrTiO$_{3}$. For SrTiO$_{3}$/LaTiO$_{3}$ interfaces annealed at high temperatures ($\sim 1000 ^{\circ}$C), which leads to Sr/La atomic interdiffusion and hence to the formation of La$_{1-x}$Sr$_x$TiO$_3$-like material, the intensity of the incoherent part was found to be dramatically reduced whereas the coherent part with a sharp Fermi cut-off is enhanced due to the spread of charge. 
These important experimental features are well reproduced by layer dynamical-mean-field-theory calculation. 
\end{abstract}


\maketitle
Strongly correlated electron systems, especially transition-metal oxides, have been the subject of numerous studies because of the variety of attractive behaviors including metal-insulator transition (MIT), spin-charge-orbital ordering and high-temperature superconductivity \cite{MIT}. 
On the other hand, heterostructures consisting of two different compounds have been extensively investigated for semiconductors because of fundamental interest in interfaces as well as of their technological importance. 
Because of the recent development in oxide thin film fabrication, atomically controlled heterostructures of transition-metal oxides have become available. 
It has therefore become important to obtain the experimental information about the electronic structures of those oxide interfaces to understand and to control the physical properties of the oxide interfaces. 
Recently, Ohtomo $et$~$al.$~\cite{Ohtomo} have made an atomic-resolution electron-energy-loss spectroscopy study of one to two layers of the Mott insulator LaTiO$_3$ (LTO) embedded in the band insulator SrTiO$_3$ (STO), and found that Ti $3d$ electrons are not completely confined within the LTO layer but are spread over the neighboring STO up to $\sim 2.5$ layers in spite of the chemically abrupt interfaces. 
Here, LTO has the $d^1$ configuration and shows antiferromagnetism below $T_N \sim 140$ K \cite{LTO-TN}, while STO has the $d^0$ configuration, i.e., the empty $d$ band. 
La$_{1-x}$Sr$_x$TiO$_3$ (LSTO), the pseudo binary alloy of LTO and STO, is a paramagnetic metal in a wide composition range of $0.08 < x \alt 1$~\cite{kumagai,LSTOphyspro}. 
Okamoto and Millis \cite{OkamotoLSTO,Okamoto-DOS} have studied the spectral function of such systems by a model Hartree-Fock and dynamical-mean-field-theory (DMFT) calculation and predicted a novel metallic behavior at the interface between the two insulators, which they call ``electronic reconstruction''. 
Popovic and Satpathy \cite{Popovic-cal} have shown by the LDA + U method that a quantum well with a wedge-shaped potential is formed in such a system and that its electronic structure can be understood in terms of the Airy-function-localized electrons distributed out of the LTO layer. 

To gain further insight into the unique electronic structure of the interfaces, information about the electronic density of states near the Fermi level ($E_F$) is essential since it is a fingerprint of the metallic behavior. 
Photoemission spectroscopy (PES) is an ideal tool to obtain such information, but the short mean-free path of photoelectron has generally prohibited access the interfacial region buried in the sample. 
In this work, we utilized the fact that STO is an $n$-type semiconductor and is ``transparent'' between $E_F$ and the O $2p$ band maximum ($\sim -3$ eV below $E_F$) and studied STO/LTO heterostructures with thin enough STO overlayers on top of the outermost LTO layers. 
We were thus able to take PES spectra of STO/LTO interfaces in various heterostructures and to gain insight into the different electronic structures at the interfaces. 

\begin{figure}
\begin{center}
\includegraphics[width=.8\linewidth]{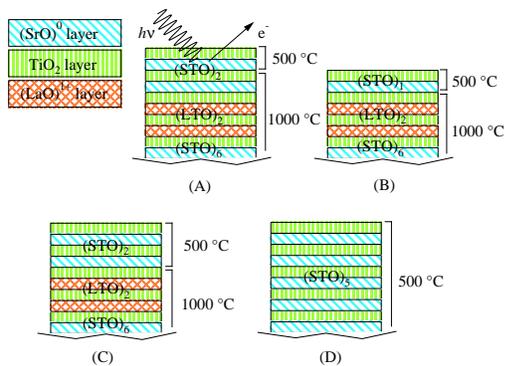}
\caption{(Color online) Schematic views of the SrTiO$_{3}$/LaTiO$_{3}$ superlattice samples (A), (B), (C), and (D). Growth temperatures are also indicated. }
\label{samples}
\end{center}
\end{figure}

We studied STO/LTO superlattice films grown on 0.05 wt\% Nb-doped STO (001)-oriented single-crystal substrates using the pulsed laser deposition (PLD) technique \cite{shibuya}. 
A Nd:YAG laser was used for heating the substrates. 
The STO substrates were wet-etched to obtain a regular step-and-terrace structure of TiO$_2$-terminated surfaces \cite{kawasaki}. 
The films were grown on the substrates at an oxygen pressure of $10^{-6}$ Torr using a KrF excimer laser ($\lambda = 248$ nm) operating at 2 Hz. 
The laser fluency to ablate the STO single-crystal and La$_2$Ti$_2$O$_7$ polycrystalline targets was $\sim 2$ J/cm$^2$. 
The film growth was monitored using real-time reflection high-energy electron diffraction (RHEED). 
Schematic views of the fabricated thin films are shown in Fig.~\ref{samples}. 
The substrate temperature was 1000$^{\circ}$C during the deposition of the [(STO)$_6$/(LTO)$_2$]$_n$ superlattices [samples (A)-(C)]. 
After 9 periods of (STO)$_6$/(LTO)$_2$, 1 layer of STO was deposited for sample (A) at the same temperature. 
The films were cooled to 500$^{\circ}$C after superlattice deposition and then another (STO)$_{n_c}$, i.e., STO cap [$n_c = 1$ for sample (A) and (B), while $n_c = 2$ for sample (C)], was deposited. 
Finally the films were annealed at an atmospheric pressure of oxygen at 500 $^{\circ}$C for 30 min to compensate for oxygen loss in the thin films and the substrates due to the high deposition temperature. 
As a reference, an STO substrate was covered with a 5 STO-layer cap [sample (D)]. 
For the thin films grown on the STO substrates, X-ray reciprocal space mapping and conventional $\theta / 2\theta$ diffraction scans were used for characterizing the samples, namely, the lattice constants, crystallinity, and grain orientation of the films were characterized. 
The surface morphology was checked by atomic force microscopy (AFM). 
The resistivity of the STO/LTO superlattices was about 300 $\mu \Omega \cdot$cm at room temperature and shows a $T^2$-dependence \cite{shibuya}. From Hall-effect measurements, the carrier density was estimated to be about $10^{21}$ cm$^{-3}$. Those transport properties are similar to those of LSTO bulk crystals \cite{R-LSTO}. 
Details of the fabrication and characterization of the films are described elsewhere \cite{shibuya}. 

Ultraviolet PES (UPS) measurements were performed using a monochromated He I line ($h\nu = 21.2$ eV) and a Scienta SES-100 electron-energy analyzer. 
Samples were transferred from the PLD chamber to the spectrometer chamber through air and no surface treatment was made prior to the PES measurements. 
The total energy resolution was set to about 20 meV. 
Resonant PES measurements in the soft x-ray region were performed at BL-2C of Photon Factory (PF), High Energy Accelerators Research Organization (KEK). The PES spectra were measured using a Scienta SES-100 analyzer. The total energy resolution was set to about 150 meV. 
All the PES measurements were made at room temperature. 
The $E_F$ position was calibrated using gold spectra. 

\begin{figure}
\begin{center}
\includegraphics[width=.8\linewidth]{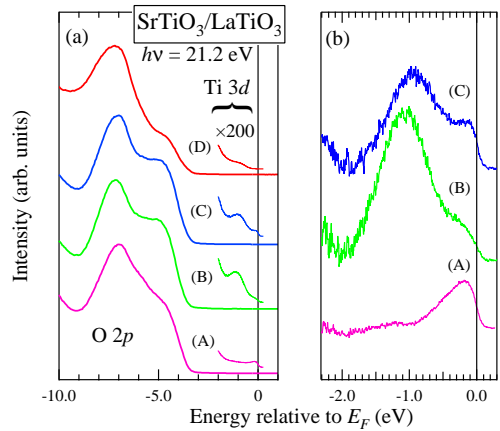}
\caption{(Color online) UPS spectra of samples (A)-(D). (a): Valence-band spectra over a wide energy range. (b): Near Fermi-level spectra after background subtraction (see text). }
\label{wide}
\end{center}
\end{figure}

The valence-band UPS spectra are shown in Fig.~\ref{wide}. 
The broad, strong feature from $-9$ to $-4$ eV [Fig.~\ref{wide} (a)] is formed by bands of primarily O $2p$ character. 
The spectra show a wide gap of about $\sim 3$ eV, consistent with the optical band gap of STO \cite{optgap}. 
Within the gap, weak emission was observed for all the samples. 
STO [sample (D)] showed the weakest emission at $\sim -1$ eV, which may be due to a small amount of oxygen defects or due to surface contaminations and may commonly exist in the spectra of the other samples since the outermost layer was STO grown at 500$^{\circ}$C in every sample.  
Therefore, we have subtracted spectrum (D) from the other spectra, resulting in Fig.~\ref{wide} (b) plotted on an enlarged scale. 
Here, the steeply rising background due to the tale of the O $2p$ band approximated by the tail of a Gaussian has also been subtracted. 
A Fermi cut-off appeared for all the STO/LTO superlattices, which may be consistent with the theoretical prediction of the ``electronic reconstruction'' between the Mott insulator and the band insulator \cite{OkamotoLSTO,Okamoto-DOS}. 
The coherent peak at $E_F$ was most clearly seen for the STO/LTO superlattice annealed at 1000$^{\circ}$C [sample (A)], which can be considered as an STO-capped La$_{1-x}$Sr$_x$TiO$_3$ film because of the La/Sr atomic interdiffusion at 1000$^{\circ}$C (Fig.~\ref{samples}). 
In the previous photoemission studies of LSTO \cite{fujimori, yoshida}, the incoherent part was clearly observed even in the high hole concentration limit of $x \agt 0.8$, where the correlation strength is thought to be negligible, and was suspected to be due to surface states. 
The weakness of the incoherent part in the present data looks more reasonable if the La/Sr intermixing indeed took place. 
Therefore, we consider that the STO cap could successfully eliminate the surface states around $-1.3$ eV \cite{STOdefect}, which overlapped the spectra of LSTO in the previous studies \cite{fujimori, yoshida}. 
Accordingly, we consider that the strong incoherent peaks seen for samples (B) and (C) are largely intrinsic and not due to surface states. 

\begin{figure}
\begin{center}
\includegraphics[width=.5\linewidth]{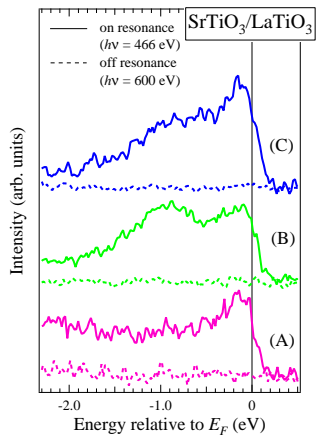}
\caption{(Color online) Ti $2p \rightarrow 3d$ resonant PES spectra taken at 466 eV. For comparison, off-resonant PES spectra taken at 600 eV are also shown. }
\label{RPES}
\end{center}
\end{figure}

In order to extract the spectra derived from the Ti $3d$ states more precisely, we performed Ti $2p \rightarrow 3d$ resonant PES, as shown in Fig.~\ref{RPES}. 
Because the emission within the band gap of STO was enhanced, these structures can be unambiguously attributed to Ti $3d$ states \cite{resonance}. 
The incoherent part became weaker in the resonant PES spectra than in the UPS spectra probably because there is stronger O $2p$ hybridization into the incoherent part than into the coherent part due to the proximity of the O $2p$ band. 
Nevertheless, both the UPS and resonant PES spectra showed similar spectral change depending on the number of top STO layers or on the growth temperature. 

\begin{figure}
\begin{center}
\includegraphics[width=.52\linewidth]{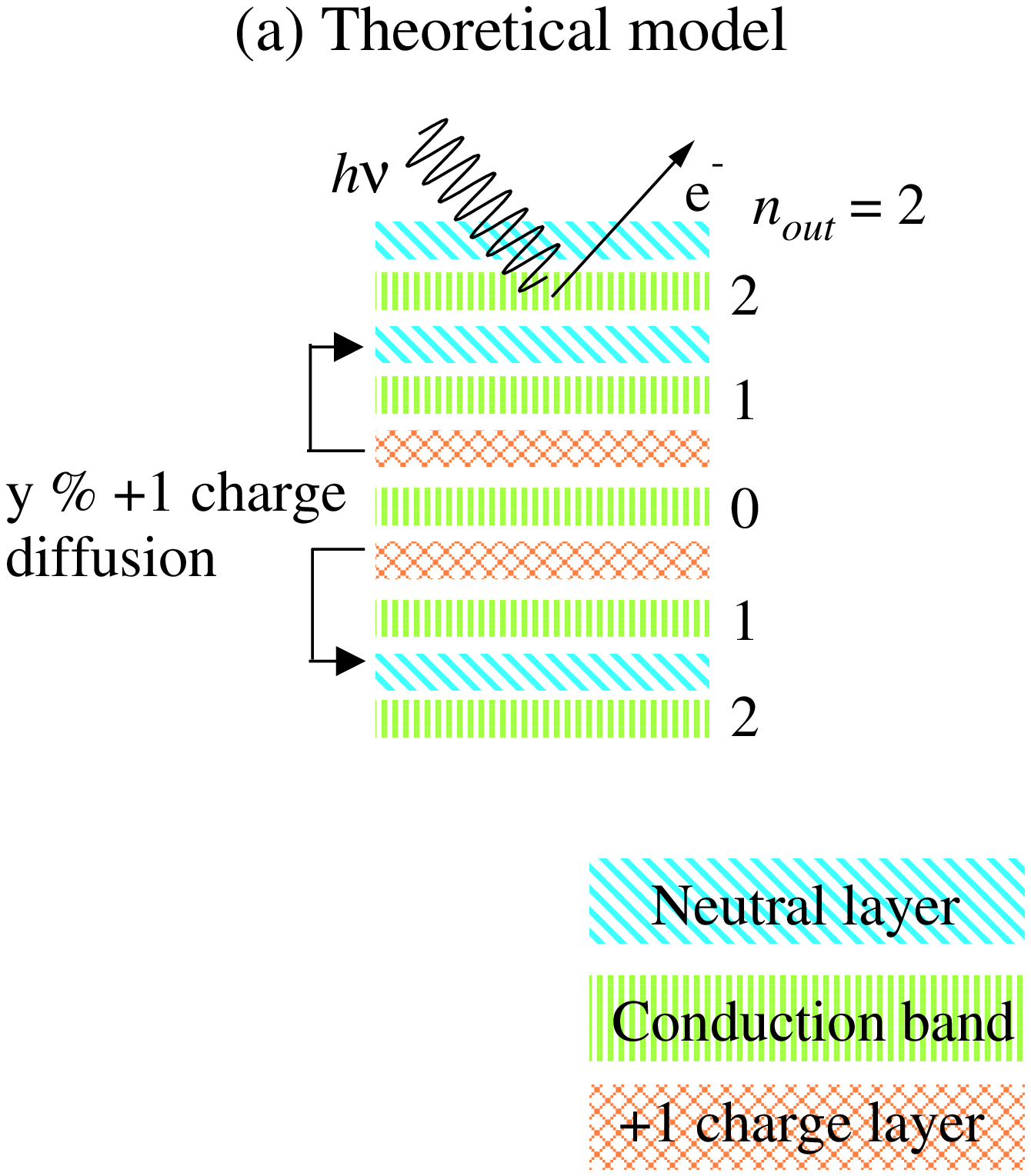}
\includegraphics[width=.43\linewidth]{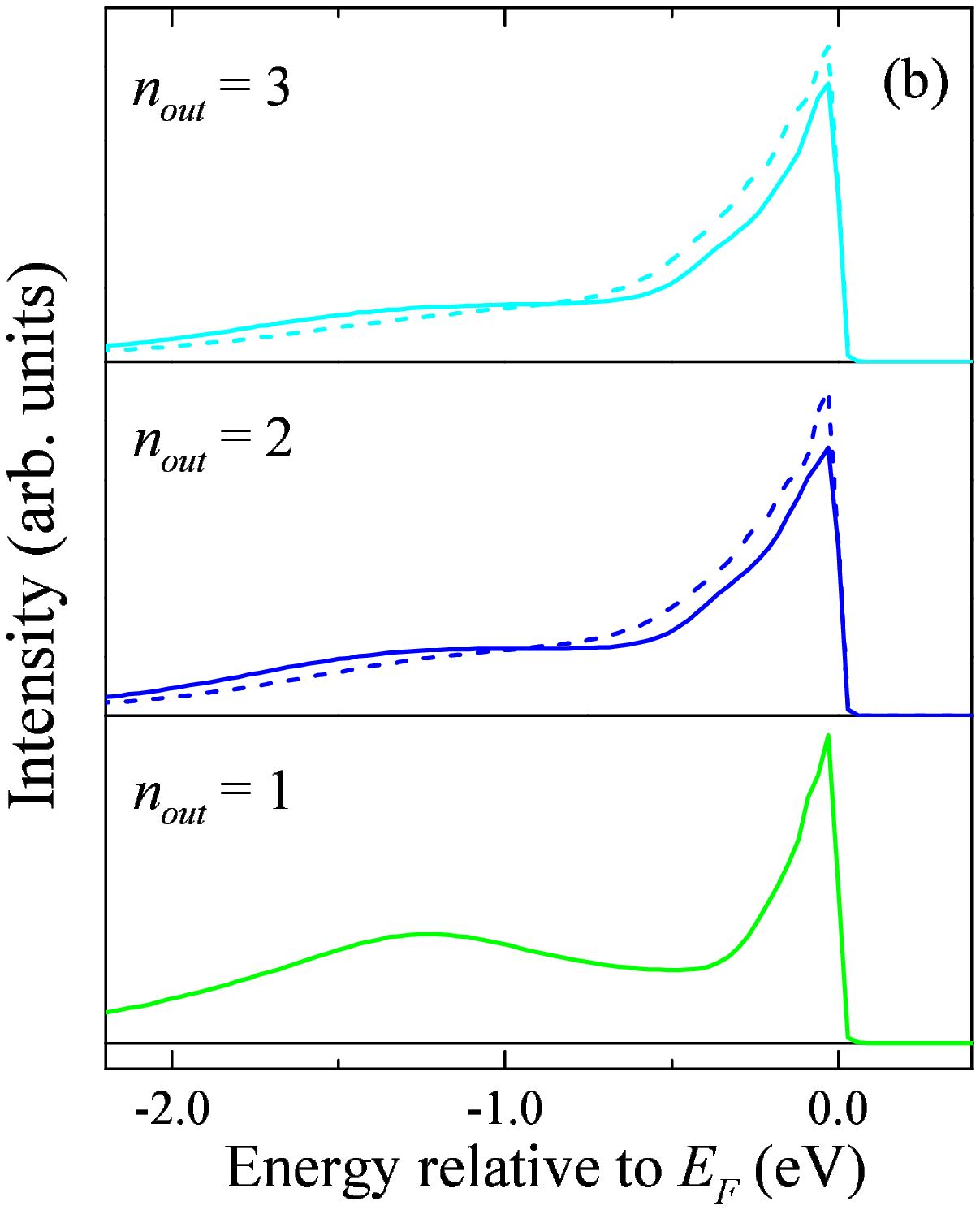}
\caption{(Color online) DMFT calculation for STO-capped STO/LTO superlattices. (a): Model for the calculation. The number of STO-cap layers (denoted by $n_{out}$) and the ratio of La/Sr mixing are varied. (b): Simulated PES spectra taking into account the effect of finite photoelectron mean-free path of three unit cells corresponding to the resonant PES. For $n_{out} = 2$ and $3$, the effect of $40 \%$ La atom diffusion is shown as broken curves. }
\label{calc}
\end{center}
\end{figure}

Now, we discuss the STO-cap thickness dependence in the spectra of the abrupt STO/LTO interfaces [sample (B) and (C)]. 
The intensity ratio of the coherent part to the incoherent part was larger for (C) than for (B). 
This difference may be explained by the spatial extent of the Ti $3d$ electrons in the interfacial region. 
According to the DMFT calculation by Okamoto and Millis \cite{Okamoto-DOS}, the intensity ratio varies with distance from the STO/LTO interface, that is, in going away from the interface into the STO side, the coherent part becomes stronger. 
Because we have measured the PES spectra from the top layers of the samples as shown in Fig.~\ref{samples}, the signals from the topmost layer is predominant due to the short photoelectron mean-free path. 
Therefore, the layer-dependent coherent-to-incoherent intensity ratio explains why the coherent part became stronger in going from sample (B) to sample (C). 

For more quantitative analysis, spectral functions have been calculated by DMFT for heterostructure models which represent the present experimental configurations. 
We adjusted the interaction $U$ to reproduce correctly the energy position of the incoherent part. 
We compare the data to spectral functions calculated by DMFT for the heterostructures shown in Fig.~\ref{calc} (a) using the model of Ref.~\cite{Okamoto-DOS} which represent the present experimental configuration. 
Note that the results of theoretical calculation practically do not change whether the ``neutral layer'' exists or not. 
We denote by $n_{out}$ the number of TiO$_2$ layers outside the LaO regime. 
We have used a generalized iteration perturbation theory method \cite{IPT} in order to solve the coupled impurity problem self-consistently. 
This method is found to give single-particle spectral functions which show better agreement with quantum Monte-Carlo results than the ones calculated using two-site DMFT. 
Details of the calculation are described in Ref.~\cite{Okamoto-DOS}. 

First, we varied the number of STO-cap layers. In comparison with Fig.~\ref{samples}, $n_{out} =2$ and $3$ corresponds to samples (B) and (C), respectively. 
Because of the finite photoemission mean-free path, PES should include contributions only from a few layers from the surface. To simulate this effect, we computed the PES spectra summing the contributions from the inner layers to be proportional to the exponential factor $e^{-r/\lambda}$, where $r$ is the distance from the surface and $\lambda$ is the photoelectron mean-free path. Here, we have assumed that the photoelectron mean-free path $\lambda$ for the resonant PES experiment is three unit cells. Resulting PES spectra are shown in Fig.~\ref{calc} (b). With the increase of $n_{out}$, the incoherent part decreases while the coherent part increases. 

Although the simulated spectra reproduce the qualitative behavior found in the experiment, the coherent-to-incoherent intensity ratio is qualitatively different between the calculation and the experiment. 
For example, the theoretical results in Fig.~\ref{calc} (b) show about twice larger weight in the coherent part than the experimental results in Fig.~\ref{RPES}. 
The large coherent weight is found in the calculations because the computed charge densities in the STO layers are less than 0.5, and for these charge densities, correlation effects are weak. 
It is possible that an extra surface potential not included in the model due for example to changes in Madelung energies and local chemistry at the TiO$_2$ termination layer confines the electrons more strongly within the LTO layers in the real interfaces. 

Next, we analyze the spectra for samples (A) and (C). 
The structure was the same for (A) and (C), however, the growth temperature was different. 
For sample (A) which was grown at 1000$^{\circ}$C, interdiffusion of the La and Sr atoms may have occurred. 
Therefore, the difference between (A) and (C) may be understood as the interdiffusion of the La and Sr atoms in sample (A). 
Again, we did DMFT calculation for the model including the atomic diffusion of La and Sr atoms [Fig.~\ref{calc} (a)]. 
The simulation of the PES spectra shown in Fig.~\ref{calc} (b) indicate that as the La and Sr atoms are intermixed, the intensity of the incoherent part decreases while that of the coherent part increases.
This tendency is consistent with the experimental spectra (A) and (C) shown in Fig.~\ref{RPES}. 
Here, the physics is an increased spreading of charges away from the LTO layer. 

According to the phase diagram proposed in Ref.~\cite{OkamotoLSTO}, the STO/LTO interface may show ferromagnetism when $U$ is large. 
Although we cannot obtain the magnetic information directly from the PES spectra, ferromagnetism may exist in STO/LTO superlattices. 
Experiments which can probe magnetism directly, such as x-ray magnetic circular dichroism measurements, are needed to investigate the magnetic behavior of the STO/LTO superlattices. 
As an extension of this work, it would be interesting to investigate the electronic structures of other superlattices, such as LaAlO$_3$/STO, which was found to be insulating or conducting depending on the termination layer \cite{OhtomoLAOSTO}.

In conclusion, we have studied the electronic structure of STO/LTO interfaces using PES. 
STO-capping successfully eliminated the surface effects and we could observe the electronic structures of the STO/LTO interfaces. 
We observed a metallic Fermi edge, indicating the formation of a metallic interface between the Mott insulator LTO and the band insulator STO. 
The spatial extent of the Ti $3d$ electrons plays an important role in explaining the different spectral features near $E_F$ depending on the STO cap layer thickness. 

The authors would like to thank K.~Ono and A.~Yagishita for their support in the experiment at PF and T.~Mizokawa for discussion. 
This work was supported by a Grant-in-Aid for Scientific Research (A16204024) from the Japan Society for the Promotion of Science and a Grant-in-Aid for Scientific Research in Priority Areas ``Invention of Anomalous Quantum Materials" from the Ministry of Education, Culture, Sports, Science and Technology. 
Two of us (MT and HW) are supported by JSPS Research Fellowships for Young Scientists. 
The work was done under the approval of Photon Factory Program Advisory Committee (Proposal No. 2003G149) at the Institute of Material Structure Science, KEK. 
SO and AJM acknowledge DOE ER 046169.


\begin{thebibliography}{10}
\bibitem{resonance}
Because the Ti $2p$ core-level photoemission excited by the second-order light of the monochromator overlapped, we took the resonant PES at the high-energy end of the $L_{2}$ edge ($h\nu = 466$ eV). The tail of the Ti $2p_{1/2}$ emission above $E_{F}$ due to the second-order light was subtracted by fitting the Ti $2p$ core-level spectra.
\end{thebibliography}
\end{document}